\documentclass[onecolumn,showpacs,pre]{revtex4}
\usepackage{graphicx}
\usepackage{subfigure}

\begin{document}

\title{Stable circulation modes in a dual-core matter-wave soliton laser }
\author{Peter Y. P. Chen}
\affiliation{School of Mechanical and Manufacturing Engineering, University of New South
Wales, Sydney 2052, Australia}
\author{Boris A. Malomed}
\affiliation{Department of Interdisciplinary Studies, School of Electrical Engineering,
Faculty of Engineering, Tel Aviv University, Tel Aviv 69978, Israel}

\begin{abstract}
We consider a model of a matter-wave laser generating a periodic array of
solitary-wave pulses. The system, a general version of which was recently
proposed in Ref. \cite{we}, is composed of two parallel tunnel-coupled
cigar-shaped traps (a reservoir and a lasing cavity), solitons being
released through a valve at one edge of the cavity. We report a stable
lasing mode accounted for by circulations of a narrow soliton in the cavity,
which generates an array of strong pulses (with $10^{3}-10^{4}$ atoms in
each, the array's \textit{duty cycle} being $\simeq 30\%$) when the soliton
periodically hits the valve.
\end{abstract}

\maketitle

\section{Introduction and the model}

One of potential applications of Bose-Einstein condensates (BECs) is using
them as a source of matter-wave (MW) laser\ beams. Atomic lasing was
predicted and demonstrated in a number of theoretical and experimental
works, starting from early ones \cite{laser-theory,laser-experiment} and
recently reviewed in \cite{review}. Especially interesting is a possibility
to design a laser operating in a \textit{soliton regime}, i.e., generating a
sequence of narrow spatial packets of coherent atom waves, as proposed, in
various settings, in Refs. \cite{soliton-laser,we}.

In a recent paper \cite{we}, we put forward a model of the soliton
MW laser based on a set of two tunnel-coupled parallel
quasi-one-dimensional (cigar-shaped) traps, one of which is a
BEC-filled reservoir, and the other plays the role of a lasing
cavity (the setup is shown in Fig. 1 of Ref. \cite{we}), with
weakly repulsive and attractive interactions between atoms in the
reservoir and cavity, respectively (the sign and size of the
respective scattering lengths can be controlled by external
magnetic field through the Feshbach-resonance effect \cite{FR};
note that static soliton states in dual-core traps, with opposite
signs of the scattering lengths in them, were studied in Ref.
\cite{Valery}). In the model, both edges of the reservoir, and the
left edge of the cavity are impenetrable, while an outlet
(\textit{valve}) at the right edge of the cavity\ releases pulses
into an outcoupling MW guide. The model was originally designed
with an intention to provide for formation of a narrow
quasi-soliton that would circulate in the cavity, bouncing from
the edges and releasing an outgoing pulse each time that it hits
the valve. The circulating soliton is supposed to replenish itself
by absorbing atoms pumped from the reservoir.

The model is based on a system of linearly coupled Gross-Pitaevskii
equations for the wave functions $\psi (x,t)$ and $\phi (x,t)$ of atoms in
the cavity and reservoir, with opposite signs in front of the nonlinear
terms. The equations are cast in a normalized form by scaling $\hbar $,
atomic mass $m$, and the nonlinearity coefficient in the cavity (in physical
units, it is $4\pi \hbar ^{2}a/m$, with $a$ the negative scattering length
of atomic collisions) to be $1$, while its counterpart, $\varepsilon $, in
the reservoir is small and positive:
\begin{eqnarray}
i\psi _{t} &=&-(1/2)\psi _{xx}-|\psi |^{2}\psi -\kappa \phi ,~  \label{psi}
\\
i\phi _{t} &=&-(1/2)\phi _{xx}+\varepsilon |\phi |^{2}\phi -\kappa \psi ,
\label{phi}
\end{eqnarray}
with $0<x<L$, i.e., $L$ is the common length of the cavity and reservoir.
Keeping $\varepsilon >0$ and the tunnel-coupling coefficient $\kappa >0$ as
free parameters, we use the residual scaling invariance of Eqs. (\ref{psi})
and \ref{phi} to fix $L\equiv 1$. The model introduced in Ref. \cite{we}
also included an external potential and $x$-dependence of the nonlinearity
coefficient in equation (\ref{psi}) for the cavity; however, the present
analysis demonstrates that stable soliton-lasing regimes can be achieved
without these ingredients.

As said above, the model assumes that both edges of the reservoir, and the
left edge of the cavity are impenetrable, the respective boundary conditions
(b.c.) being $\phi (x=0)=\phi (x=1)=\psi (x=0)=0$. The remaining b.c.,
corresponding to the valve at the cavity's right edge ($x=1$) is
\begin{equation}
\psi _{x}(x=1)=iq\psi (x=1),  \label{q}
\end{equation}
with $q>0$ (as explained in Ref. \cite{we}, it corresponds to the
potential drop $\Delta U=q^{2}/2$ at the outcoupling edge). After
straightforward manipulations with Eqs. (\ref{psi}) and
(\ref{phi}), this b.c. yields a balance equation for the total
norm of the wave functions $\psi $ and $\phi $, which is
proportional to the net number of atoms in the traps,
\begin{equation}
N=\int_{0}^{1}\left( |\psi (x)|^{2}+|\phi (x)|^{2}\right) dx\equiv
N_{1}+N_{2}.  \label{N}
\end{equation}
The balance equation can be cast in the following compact form:
\begin{equation}
R(t)\equiv -\frac{dN}{dt}=q\left\vert \psi (x=1,t)\right\vert ^{2}.
\label{E}
\end{equation}
Note that the valve gets shut in both limits of $q=0$ and
$q\rightarrow \infty $ [in the latter case, Eq. (\ref{q}) shows
that $\left\vert \psi (x=1)\right\vert ^{2}\sim q^{-2}$, hence $R$
vanishes as $1/q$]. Monitoring $R(t)$ by means of Eq. (\ref{E}),
one can characterize the lasing regime (see below). The model does
not include an explicit form of the outcoupling waveguide (at
$x>1$), which shapes the released pulses into solitons, as soliton
formation in a uniform waveguide is a well-studied problem.

The system is controlled by five parameters: $\varepsilon $,
$\kappa $, $q$, and initial numbers of atoms, $N_{1}^{(0)}$ and
$N_{2}^{(0)}$, in the cavity and reservoir, respectively. It is,
of course, difficult to exhaustively explore the corresponding
parameter space, looking for physically meaningful regimes. In
particular, a lasing mode based on circulation of a well-defined
soliton in the cavity was found in Ref. \cite{we} in some
parameter region, but it was unstable: after no more than $40$
circulations (usually fewer), the soliton would quickly come to a
halt, getting much broader. Instead of the unstable circulation
mode, a stable regime of periodic release of pulses was found in a
large part of the section of the parameter space explored in Ref.
\cite{we}. In that regime, a broad ``lump" stays put in the
cavity, performing cycles of stretching and compression. When its
right wing periodically reaches the outlet (valve), weak pulses
are released.

The mode of periodic vibrations of the immobile lump is not a spectacular
one, and, in practical terms, its drawback is that each pulse generated this
way contains a small number of atoms, typically not much larger than $10$
\cite{we}. It would be more interesting to find a mode of stable soliton
circulations in the cavity. In this paper, we demonstrate in direct
simulations that exploration of a broader parameter region reveals a stable
soliton-circulation mode. The existence and stability of this mode are
directly explained by introducing a simplified analytically tractable
version of the model, in which we apply the perturbation theory to the
soliton circulating in the cavity. The number of atoms in each released
pulse is $N_{\mathrm{pulse}}\sim 10^{3}-10^{4}$, which is quite sufficient
to shape the pulses into true MW solitons ($N_{\mathrm{pulse}}$ is larger,
roughly, by a factor of $100$ than in was provided by the above-mentioned
pulsation mode in Ref. \cite{we}). The quality of the soliton array is
characterized by its \textit{duty cycle }(the ratio of the temporal width of
each pulse to the duration of the cycle), $\simeq 30\%$, which makes it
possible to avoid the unwanted effect known (in terms of telecommunications
\cite{Agrawal}) as \textit{inter-symbol interference }(conspicuous overlap
between pulses through their extended tails); in the regime investigated in
Ref. \cite{we}, the duty cycle was $\simeq 50\%$.

The rest of the paper is organized as follows. In Section II, we
display generic examples of two different stable
intracavity-circulation regimes, periodic and quasi-periodic ones,
found in direct simulations of Eqs. (\ref{psi}), (\ref{psi}),
which provide for the MW lasing in the soliton format (we also
display an example of another generic regime possible in the
system, \textit{viz}., persistent irregular circulations). In\
Section III, we summarize results of the systematic analysis in
the form of stability charts for the periodic and quasi-periodic
circulation modes in a relevant parameter space, and present
dependences of essential characteristics of the modes on
parameters of the system. The simplified version of the model and
its analytical perturbative solutions, that demonstrates the
existence of the stable-circulation regime, are presented in
Section IV. The paper is concluded by Section V, where, in
particular, we give estimates for physical characteristics of the
proposed pulsed MW laser.

\section{Stable circulation modes}

Numerical simulations of Eqs. (\ref{psi}) and (\ref{phi}) were performed by
means of a finite-element pseudospectral method. In most cases, initial
configurations were built (following Ref. \cite{we}) as stationary solutions
to the same equations, but with b.c. (\ref{q}) replaced by $\psi (x=1)=0$.

\begin{figure}[tbp]
\subfigure[]{\includegraphics[width=2.5in]{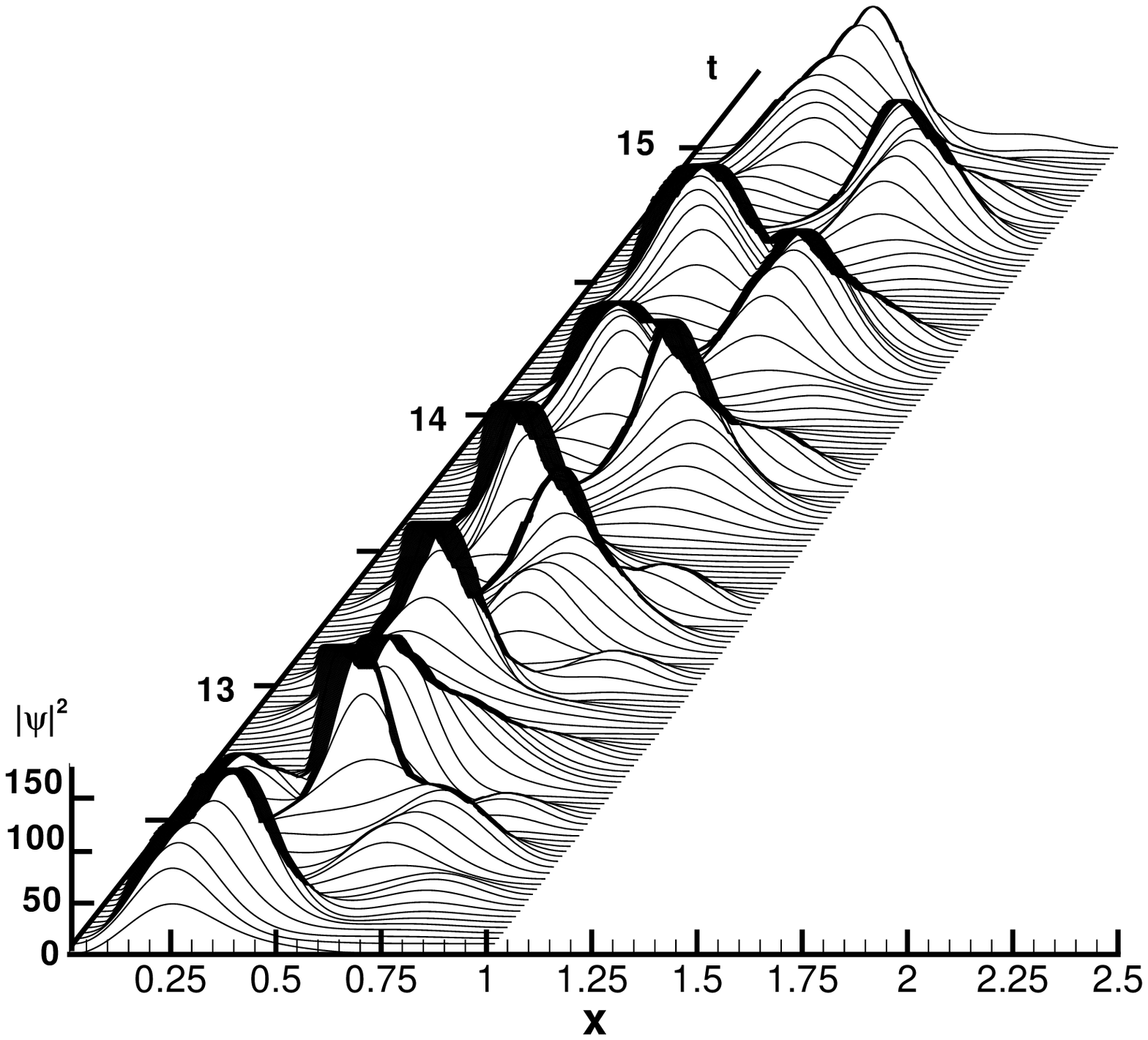}}
\subfigure[]{\includegraphics[width=2.5in]{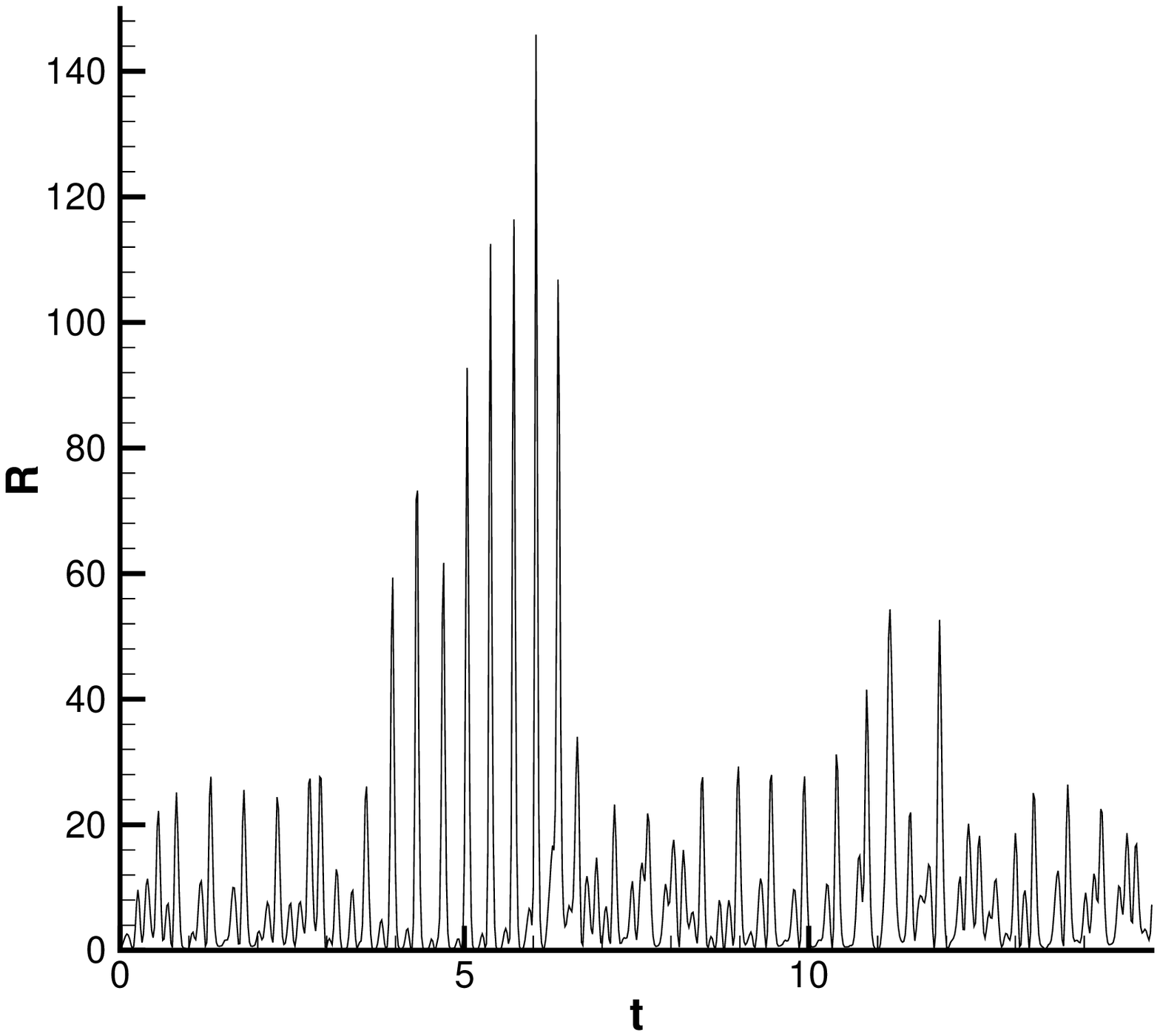}} \caption{A
typical example of the spatiotemporal density distribution in the
cavity, $\left\vert \protect\psi (x,t)\right\vert ^{2}$ (a), and
the corresponding release rate, defined by Eq. (\protect\ref{E}),
as a function of time (b), in the case of persistent irregular
circulations of the intra-cavity soliton circulations. Parameters
are $\protect\kappa =0.8$, $\protect\varepsilon =0.0025$,
$q=32.5$, and initial values of the norm of the wave function in
the cavity and reservoir are $N_{1}^{(0)}=0$ and
$N_{2}^{(0)}=1000$.} \label{fig0}
\end{figure}

Exploring the parameter space of the model in a more systematic
way than in Ref. \cite{we}, we were able to find three distinct
types of regimes featuring \emph{persistent circulations} of a
narrow soliton in the cavity and, accordingly, the MW lasing in
the pulsed format. The first type is persistent in the sense that
the soliton keeps to circulate as long as the depletion of the
reservoir does not become conspicuous, but the circulations are
irregular (quasi-random). An example of this regime is shown in
Fig. \ref{fig0}, in terms of the evolution of the density profile
in the cavity, $|\psi (x,t)|^{2}$, and the corresponding release
rate, $R(t)$. We do not report more details about this mode, as it
seems unusable for applications.

\begin{figure}[tbp]
\subfigure[]{\includegraphics[width=2.5in]{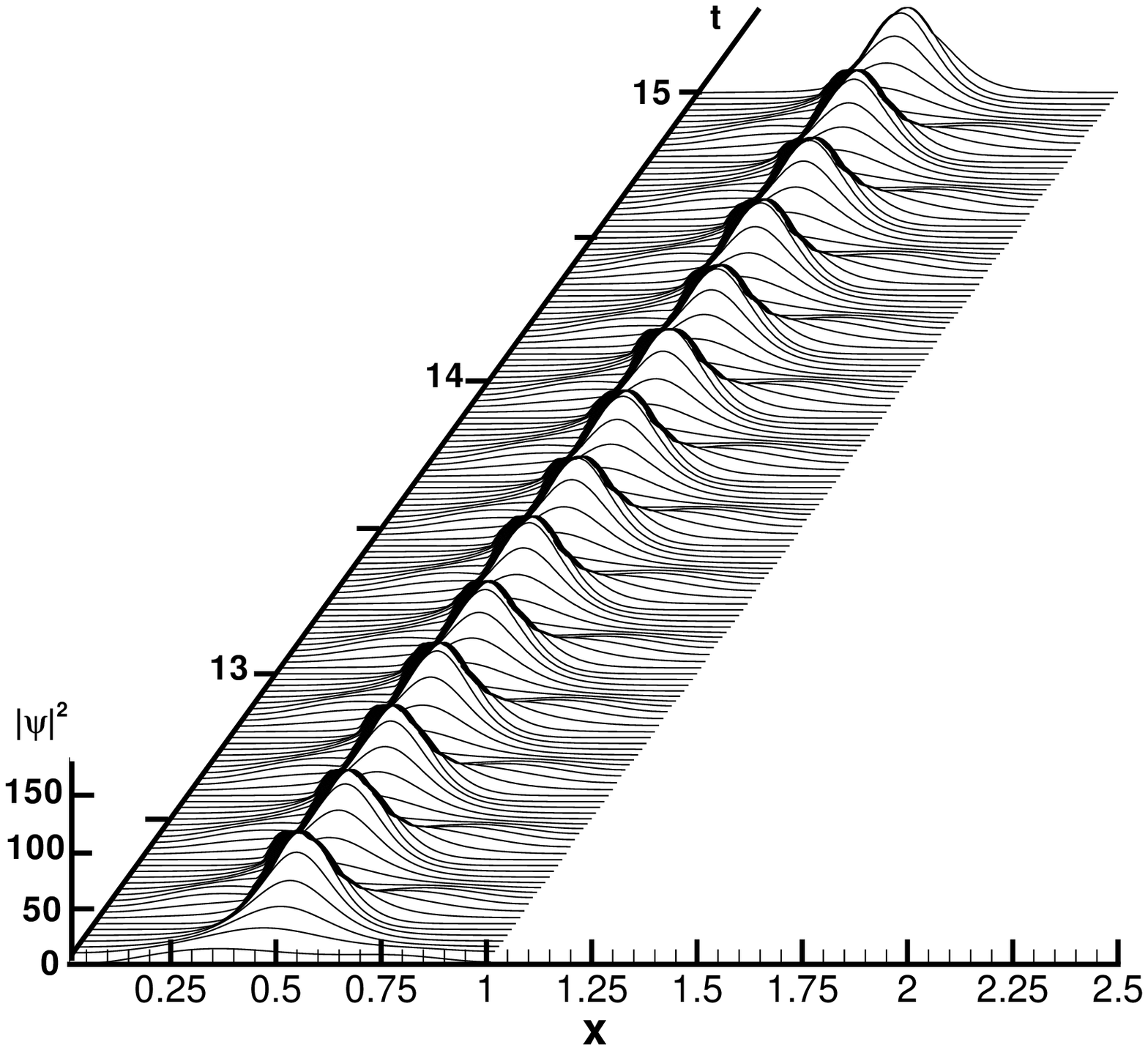}}
\subfigure[]{\includegraphics[width=2.5in]{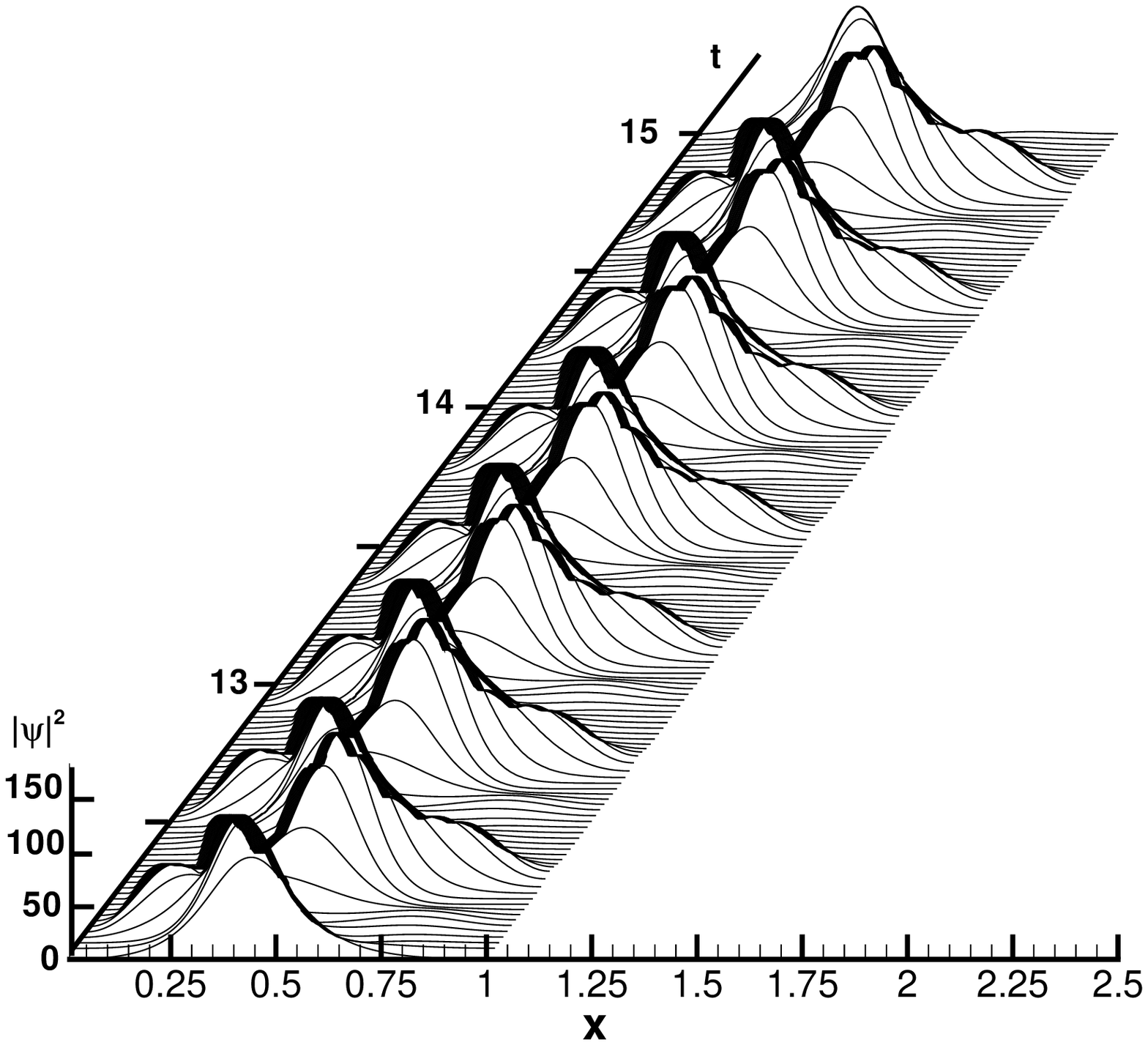}}
\caption{Generic examples of the spatiotemporal density
distribution in the cavity, $\left\vert \protect\psi
(x,t)\right\vert ^{2}$, in regimes of periodic (a) and
quasiperiodic (b) intra-cavity soliton circulations, for
$N_{1}^{(0)}=10$, $N_{2}^{(0)}=1000$, $\protect\kappa =0.8$,
$\protect\varepsilon =0.002$, and (a) $q=12.5$, (b) $q=21.0$.}
\label{fig1}
\end{figure}

\begin{figure}[tbp]
\subfigure[]{\includegraphics[width=2.5in]{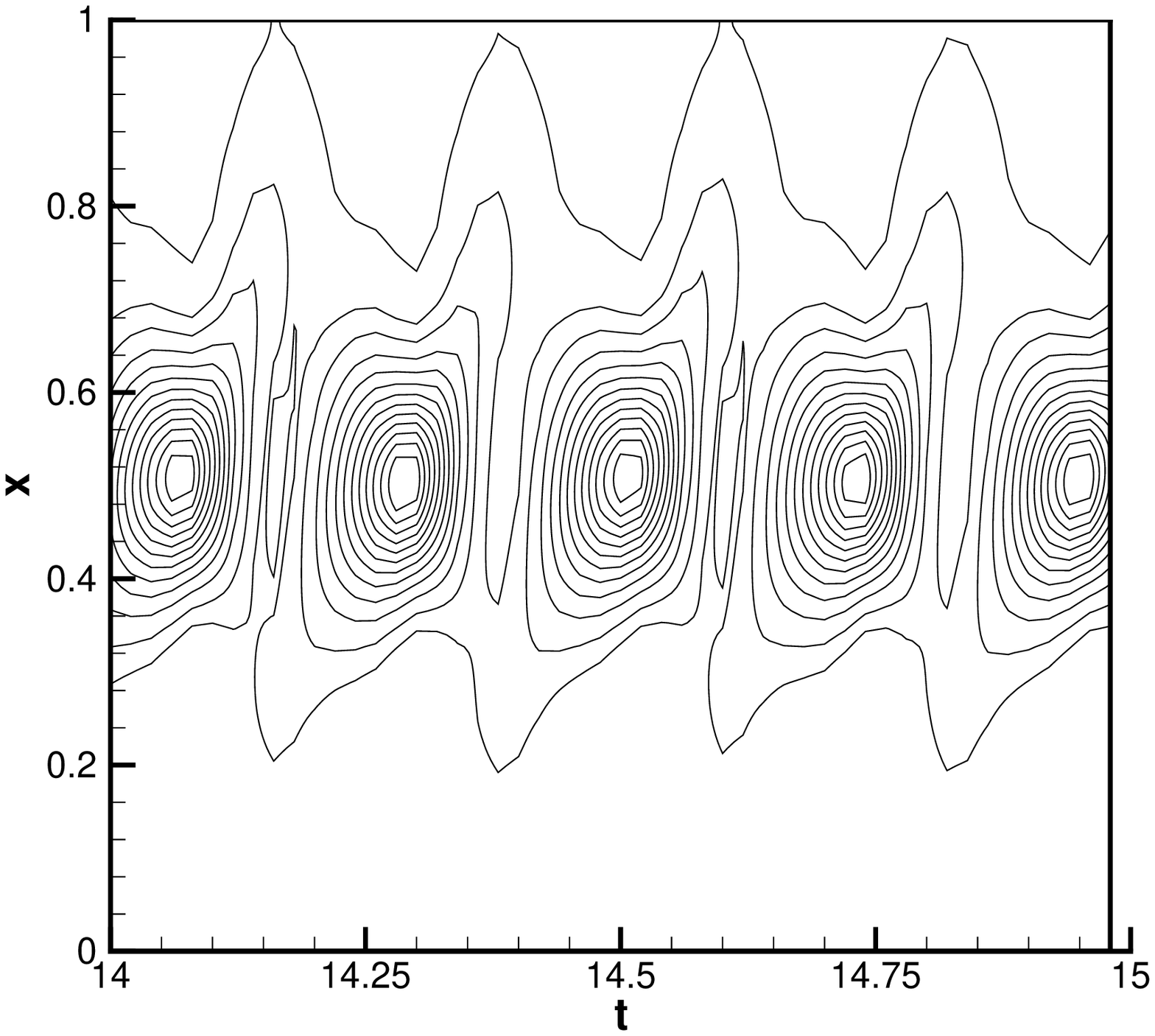}}
\subfigure[]{\includegraphics[width=2.5in]{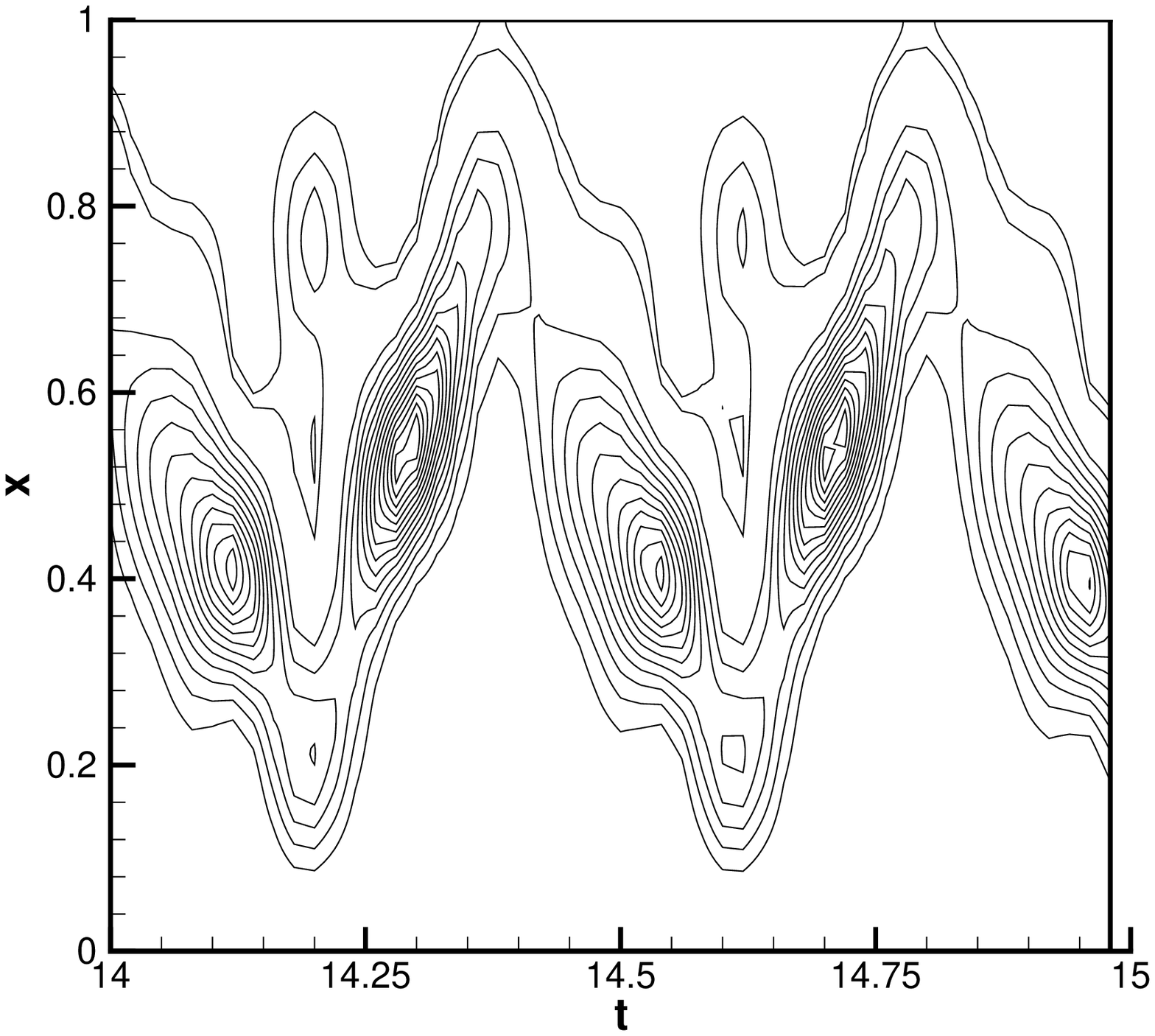}}
\caption{The same as in Fig. \protect\ref{fig1}, shown by means of
contour plots for the spatiotemporal field of $|\protect\psi
(x,t)|^{2}$.} \label{fig2}
\end{figure}

\begin{figure}[tbp]
\subfigure[]{\includegraphics[width=2.5in]{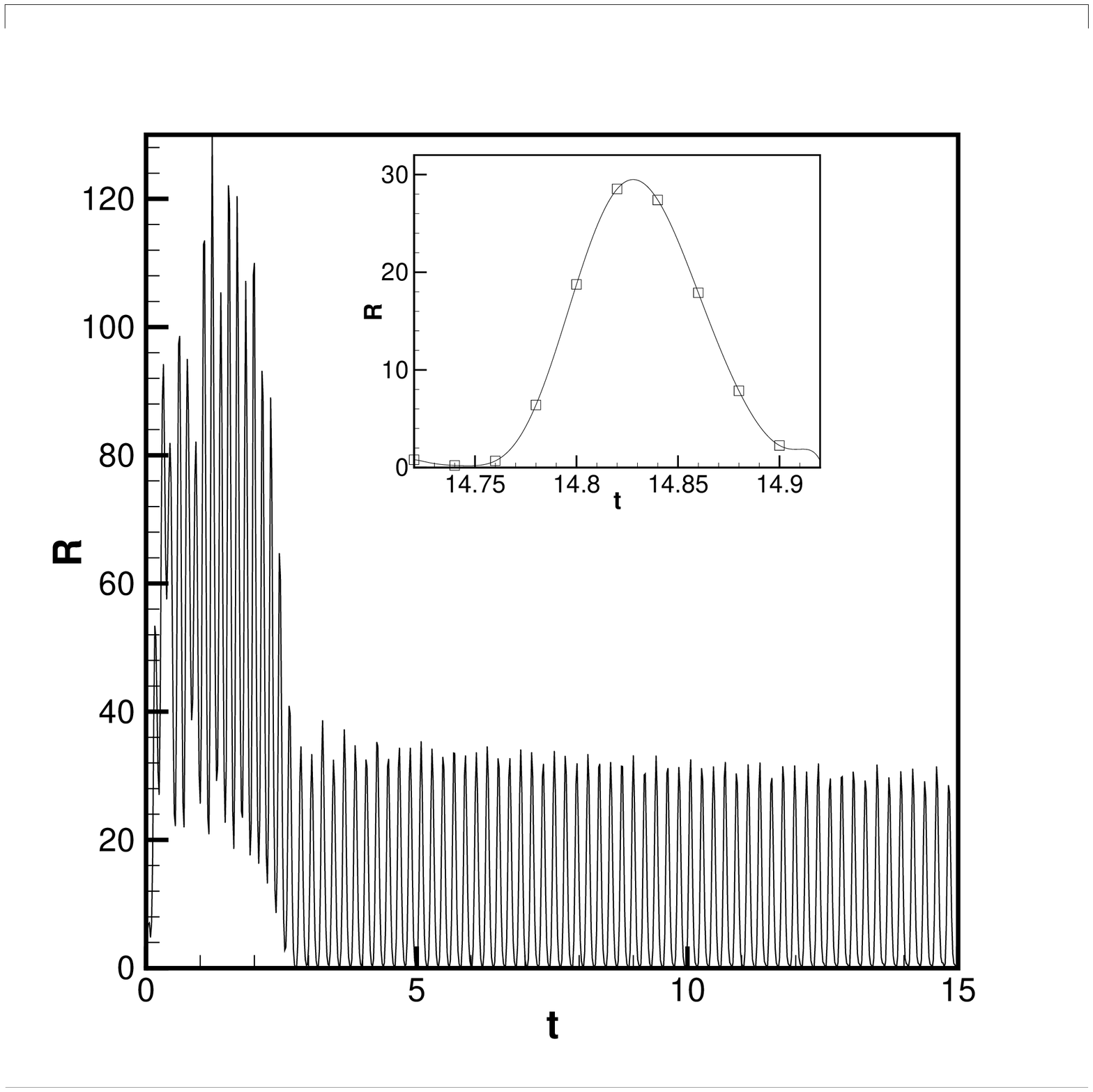}}
\subfigure[]{\includegraphics[width=2.5in]{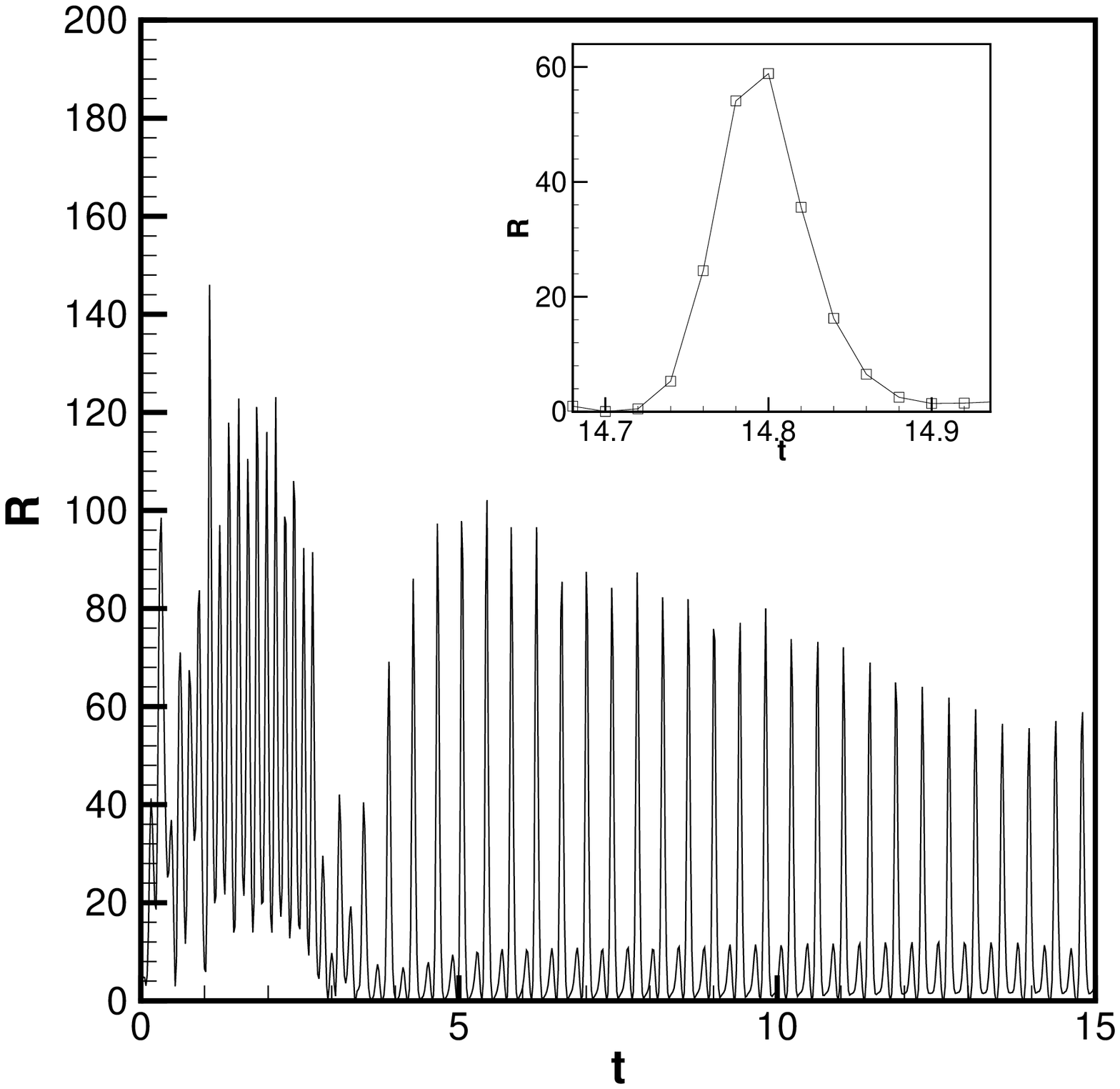}}
\caption{The release rate $R$, defined as per Eq.
(\protect\ref{E}), vs. time, in the same cases as in Figs.
\protect\ref{fig1} and \protect\ref{fig2}. Insets in each panel
show the shape of individual pulses in the established regime,
with the continuous lines being a guide to the eye. The
\textit{duty cycle} of the (quasi-) periodic pulse arrays, defined
as the ratio of the full time width of the temporal pulse at its
half-maximum to the full period, is $32\%$ (a) and $27\%$ (b).}
\label{fig3}
\end{figure}

More promising are two other varieties of the circulation mode,
one periodic and another one quasi-periodic. Typical examples of
spatiotemporal patterns specific to these modes are displayed in
Figs. \ref{fig1} and \ref{fig2}, by means of both spatiotemporal
profiles of $|\psi (x,t)|^{2}$ and respective contour plots. The
corresponding temporal patterns of $R(t)$, which determine the
profile of the soliton streams released from the cavity (the same
way as the temporal shape of an input optical signal coupled into
a nonlinear fiber determines the form of the soliton array created
in it \cite{Agrawal}), are displayed in Fig. \ref{fig3}. It is
seen that both modes, periodic and quasi-periodic ones, readily
establish themselves after a transient (this self-establishment
was adopted as a criterion of the mode's stability). In the former
case, the soliton circulates in the central part of the cavity,
while in the quasi-periodic mode it sweeps the entire domain and
strongly hits the valve, in an almost periodic fashion; for this
reason, the amplitude of the generated pulses in essentially
higher in the latter case. Another notable difference of the
quasi-periodic regime from its periodic counterpart is
\textit{period doubling} obvious in Fig. \ref{fig2}(b), due to
which each cycle of the quasi-periodic oscillations is composed of
a narrow tall pulse to which a small hump is attached.

\begin{figure}[tbp]
\subfigure[]{\includegraphics[width=2.5in]{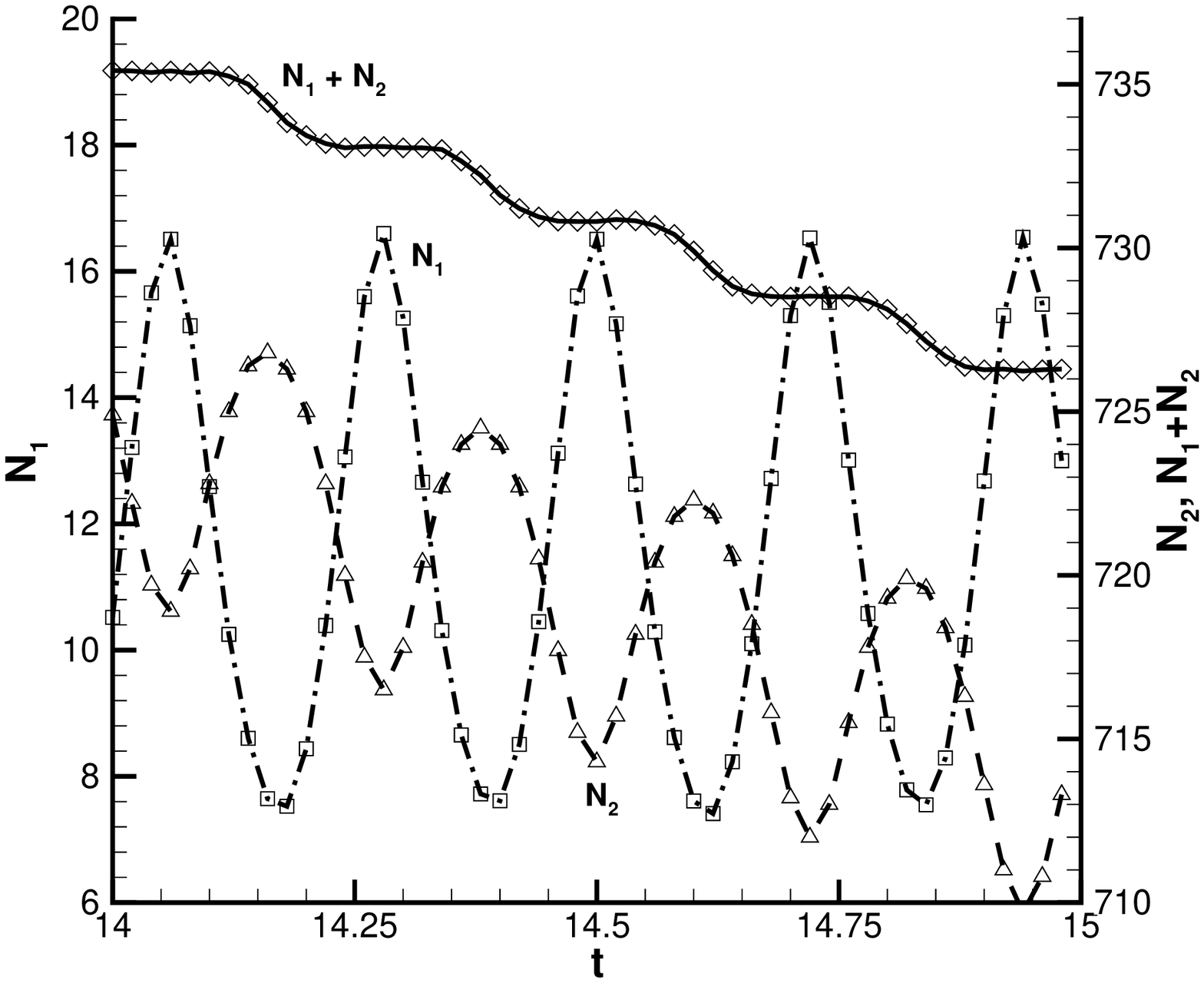}}
\subfigure[]{\includegraphics[width=2.5in]{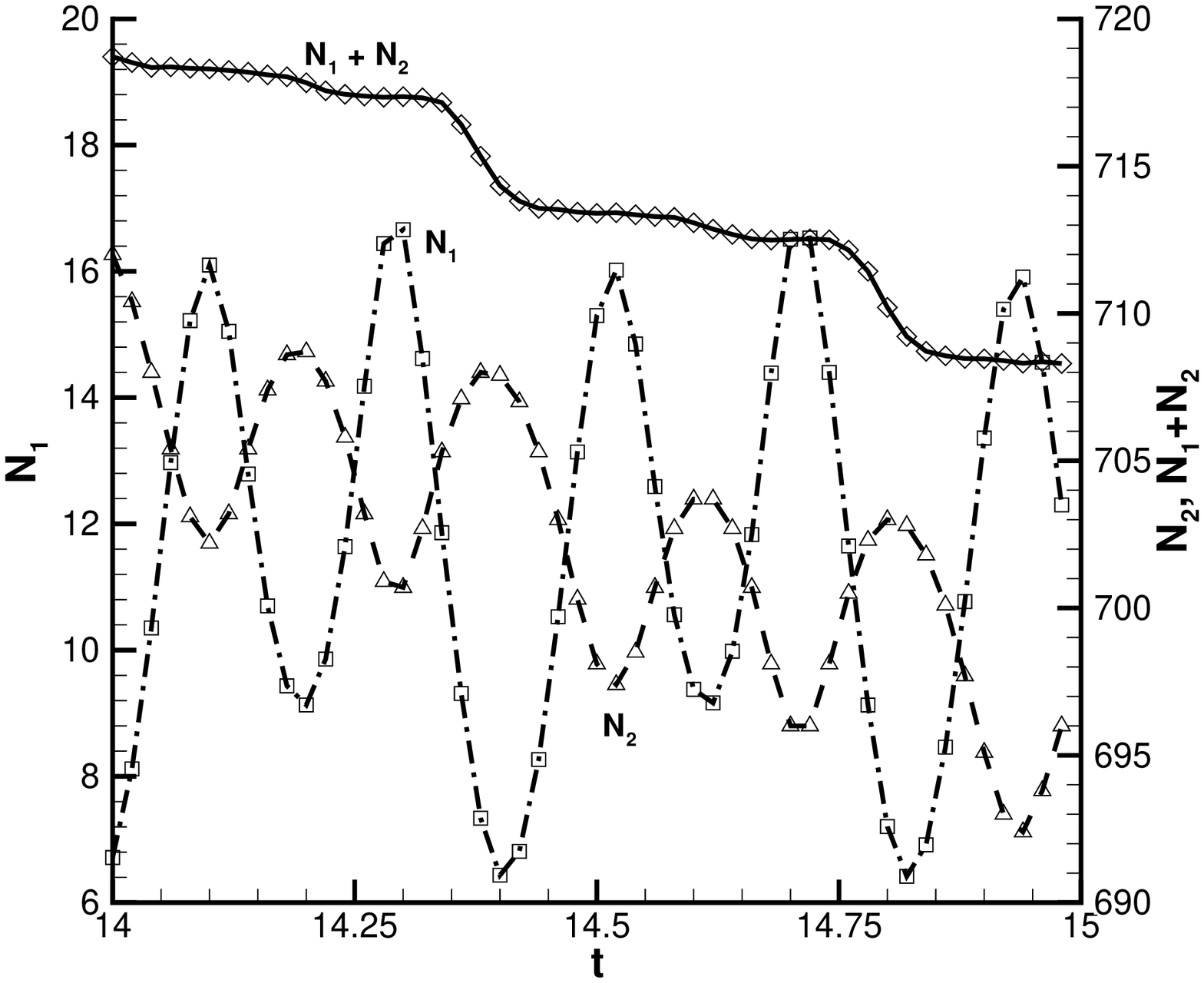}}
\caption{The evolution of the total norm, $N_{1}+N_{2}$
(rhombuses), and norms in the lasing cavity and reservoir, $N_{1}$
(squares) and $N_{2}$ (triangles), in the same established
dynamical regimes [periodic and quasiperiodic ones, (a) and (b)]
which are shown in Figs. \protect\ref{fig1} and
\protect\ref{fig2}.} \label{fig1n}
\end{figure}

An additional relevant characteristic of the periodic and
quasi-periodic regimes is given in terms of the time dependence of
norms $N_{1}$ and $N_{2}$ in the cavity and reservoir, as well as
the total norm, $N_{1}+N_{2}$ [see Eq. (\ref{N})]. For the same
cases as shown in Figs. \ref{fig1} and \ref{fig2}, the dependences
are displayed in Fig. \ref{fig1n}. They clearly highlight the
character of the dynamical regimes: a relatively fast matter
exchange between the reservoir and cavity, and slower depletion of
the total norm, which actually happens by jumps when a new pulse
is released into the outcoupling MW waveguide.

All stable periodic and quasi-periodic soliton-lasing modes found
at other values of parameters (see stability charts presented in
the next section) are very similar to the examples displayed in
Figs. \ref{fig1} - \ref{fig1n}. As concerns the persistent
irregular regimes, they are quite similar to the example displayed
in Fig. \ref{fig0}).

\section{Stability maps for the circulation modes.}

\begin{figure}[tbp]
\subfigure[]{\includegraphics[width=2.5in]{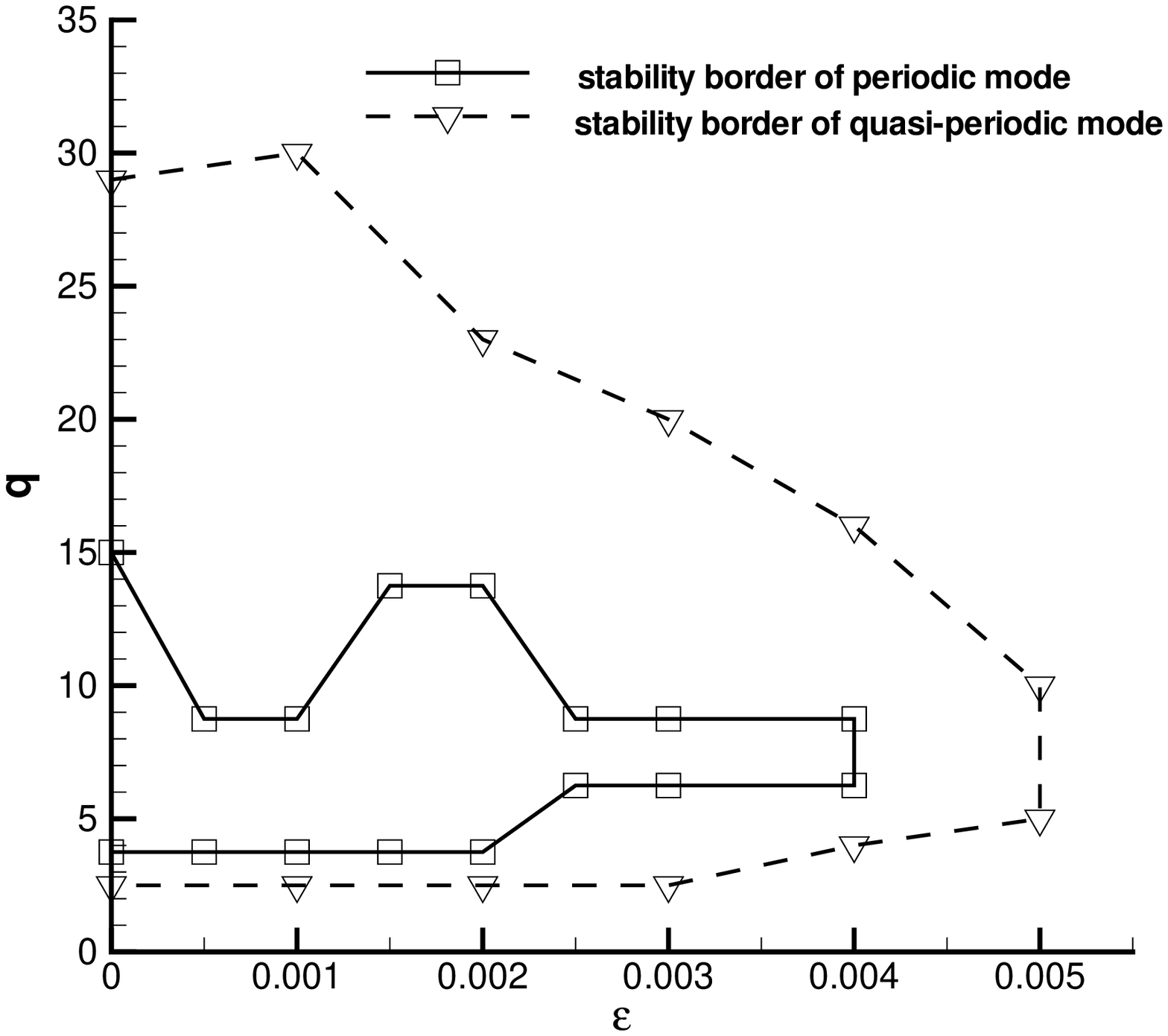}}
\subfigure[]{\includegraphics[width=2.5in]{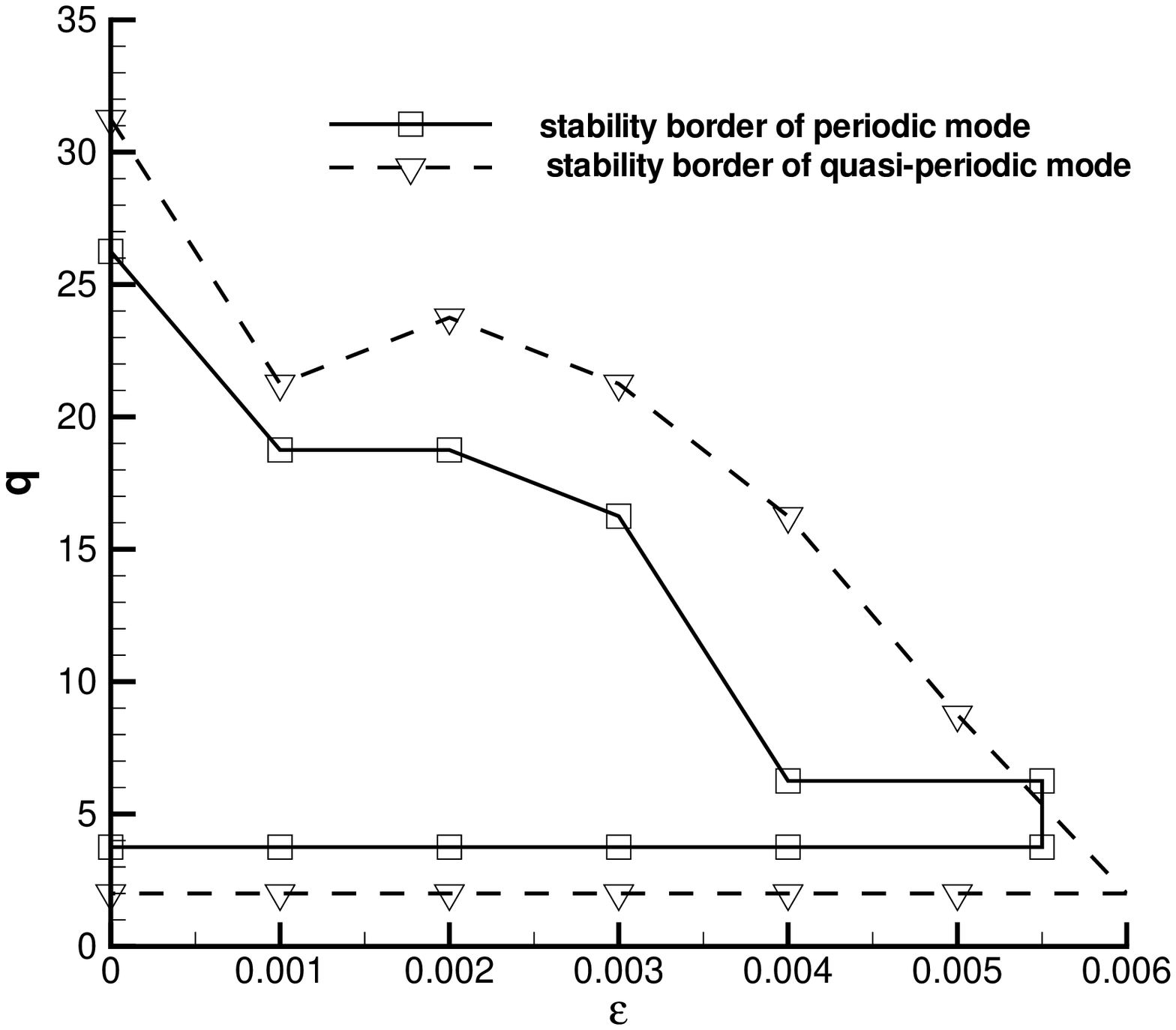}}
\caption{Stability borders of the periodic and quasi-periodic
regimes in the parameter plane $\left( \protect\varepsilon
,q\right) $, for $\protect\kappa =0.8$, $N_{2}^{(0)}=1000$, and
(a) $N_{1}^{(0)}=0$, (b) $N_{1}^{(0)}=10$.} \label{fig4}
\end{figure}

\begin{figure}[tbp]
\includegraphics[width=7.0cm,height=4.5cm]{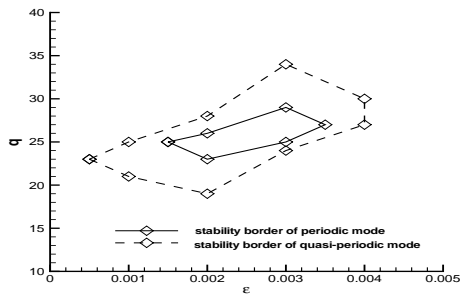}
\caption{The same as in Fig. 4(b), except that $\protect\kappa =0.5$.}
\label{fig5}
\end{figure}

As the model contains several parameters, it is necessary to identify those
which strongly affect the stability of the lasing modes. For this purpose,
we summarize results of a large number of simulations in charts displayed in
Figs. \ref{fig4} and \ref{fig5}. The periodic and quasi-periodic circulation
modes are stable, respectively, inside the area bounded by the continuous
line, and between dashed and continuous borders. The charts displayed in
these figures were obtained by varying $q$ and $\varepsilon $ in steps of $2$
and $0.001$, respectively, unless strong sensitivity to variation of the
parameters dictated to use reduced steps.

Comparison of Figs. \ref{fig4}(a) and \ref{fig4}(b), which differ
by values of the initial number of atoms in the cavity,
demonstrates that the outer stability border (appertaining to the
quasi-periodic regime) is relatively insensitive to this
parameter, while the inner border, which separates the stability
areas of the periodic and quasi-periodic regimes, is conspicuously
affected by the initial conditions. The effect of variation of the
strength of the linear coupling between the reservoir and cavity,
$\kappa $, on the stability region is illustrated by comparison of
Figs. \ref{fig4}(b) and \ref{fig5}. Naturally, the decrease of
$\kappa $ entails shrinkage of the stability area, as it becomes
harder for the circulating soliton to replenish itself by
collecting matter pumped from the reservoir (see also the
analytical model presented in Section V). In addition, the results
suggest that, at smaller $\kappa $, the stability region shifts to
higher values of $q$. Detailed analysis shows that the increase of
$q$ makes the release rate $R$ smaller in this case, thus
compensating the decrease of the pumping rate due to the reduction
in the coupling strength.

A noteworthy peculiarity of Fig. \ref{fig4} is a nearly flat lower stability
border, which is found at $q=q_{\min }\approx 4$. An explanation for this
flat border is provided by the analytical model presented below.

Beneath lower borders of the stability areas in Figs. 4 and 5, no
circulation regime is possible. Above the upper stability borders,
persistent irregular circulations take place, similar to the example shown
in Fig. \ref{fig1}. Outside the stability area, there is an additional
border between the no-circulation and irregular-circulation regimes, which
we do not display here, as it is of little relevance to applications.

\begin{figure}[tbp]
\includegraphics[width=7.0cm,height=4.5cm]{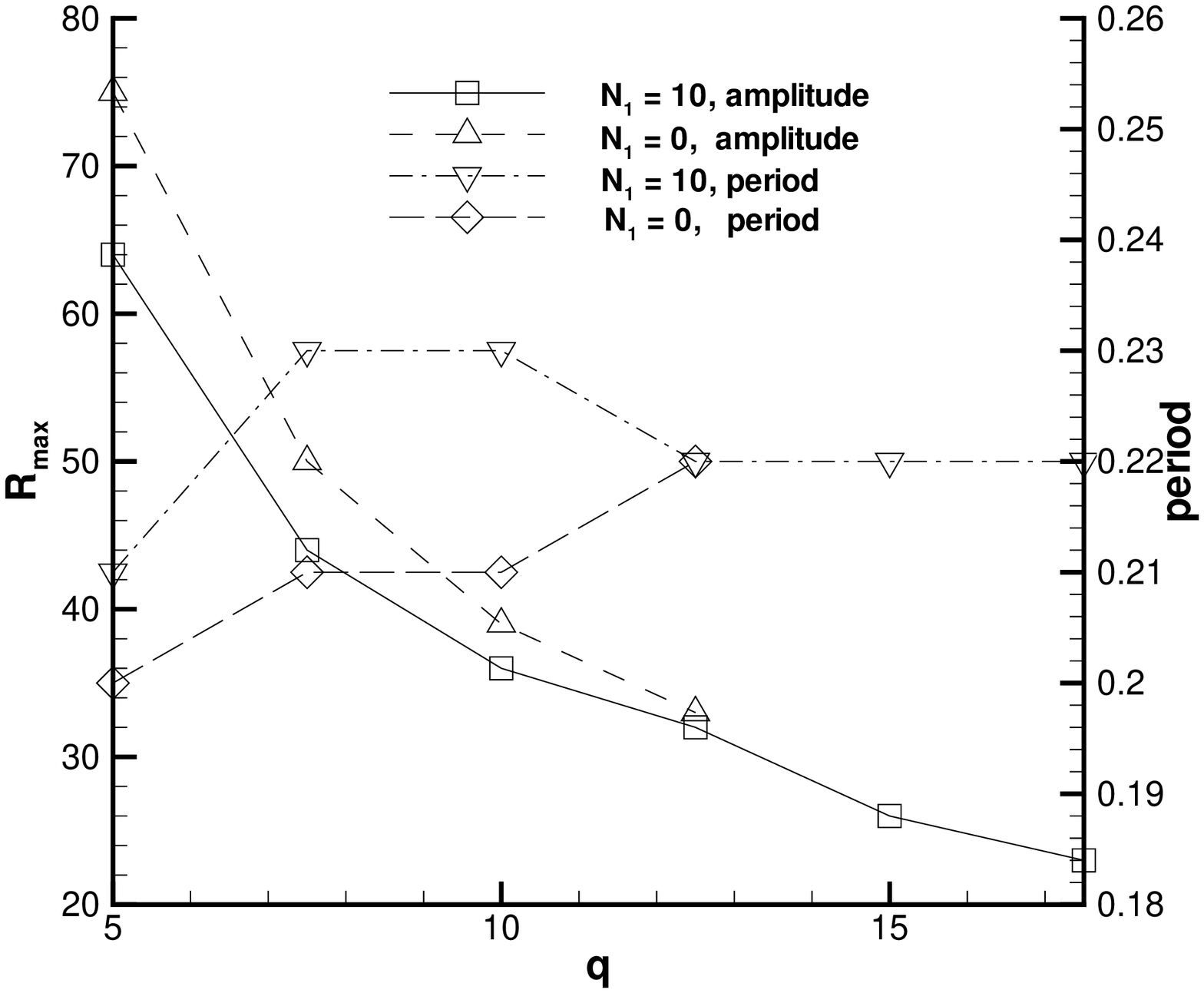}
\caption{The maximum value (``amplitude") of the release rate $R$,
and its oscillation period, versus the outcoupling coefficient
$q$, for $\protect\kappa =0.8$, $\protect\varepsilon =0.002$,
$N_{2}^{(0)}=1000$, and two different values of $N_{1}^{(0)}$.
Data sets for $N_{1}^{(0)}=0$ terminate at points beyond which
stable circulation modes are not found.} \label{fig6}
\end{figure}

Effects of the variation of control parameters on characteristics
of the generated pulse arrays were investigated too. In
particular, a change in $N_{2}^{(0)}$ does not affect the results
in any tangible way, and arrays of the outcoupling pulses are
quite insensitive to the variation of $\varepsilon $ too. On the
other hand, the dependence on the valve parameter $q$ is
conspicuous, as seen in Fig. \ref{fig6}.

\section{The analytical model}

A qualitative explanation to the stable regime of periodic circulation of
the soliton in the cavity may be provided by the perturbation theory, which,
at the zero order, assumes shuttle oscillations of a soliton in the cavity
of length $L$ with zero b.c., $\psi (x=0)=\psi (x=L)=0$, with no coupling to
the reservoir, $\kappa =0$. The respective zero-order approximation for the
soliton is
\begin{eqnarray}
\psi _{0}(x,t) &=&\eta ~\mathrm{sech}\left( \eta (x-\xi (t)\right)
~e^{ivx+i\Phi (x)},~  \nonumber \\
\frac{d\xi }{dt} &=&v,~\frac{d\Phi }{dt}=-\frac{1}{2}\left( v^{2}-\eta
^{2}\right) ,  \label{soliton}
\end{eqnarray}
where $\eta $ and $v$ are its amplitude and instantaneous velocity. In this
approximation, Eq. (\ref{psi}) conserves two dynamical invariants,
\textit{viz}., the norm, $N_{1}=\int_{0}^{L}\left\vert \psi
(x)\right\vert ^{2}dx$, see Eq. (\ref{N}) (in this section, we do
not set $L=1$, to present the dependence of the results on
parameters in a clearer form), and the energy (Hamiltonian),
\begin{equation}
H_{1}=\frac{1}{2}\int_{0}^{L}\left( \left\vert \psi _{x}\right\vert
^{2}-|\psi |^{4}\right) dx.  \label{H}
\end{equation}
Note that, for unperturbed soliton (\ref{soliton}), the values of these
invariants are
\begin{equation}
\left( N_{1}\right) _{\mathrm{sol}}=2\eta ,~\left( H_{1}\right)
_{\mathrm{sol}}=\eta v^{2}-\frac{1}{3}\eta ^{3}.  \label{NH}
\end{equation}

Zero b.c. imply the presence of formal mirror images of the soliton beyond
the points of $x=0$ and $x=L$, i.e., $\psi (-x)=-\psi (x)$, and $\psi
(x)=-\psi (2L-x)$, respectively. As is well known, a pair of solitons with
opposite signs repel each other \cite{KM}, which explains bounces of the
soliton from edges of the cavity and its shuttle motion.

In the first nontrivial approximation, we treat the coupling
coefficient $\kappa $ and b.c. coefficient $q$ in Eq. (\ref{q}) as
small parameters. To predict the possibility of stable periodic
circulations of the soliton in this case, we will treat it as a
quasiparticle trapped in an infinitely deep potential box, between
$x=0$ and $x=L$. Then, the shuttle motion can be predicted by
means of balance equations for the norm and energy [this approach
was elaborated in detail for a kink (topological solitons) moving
as a shuttle in models based on a perturbed sine-Gordon equation
\cite{KM}].

First, note that Eq. (\ref{E}) makes it possible to calculate the loss of
norm $N_{1}$ for the soliton bouncing from the right edge ($x=L$):
\begin{equation}
\left( \Delta N_{1}\right) _{q}=-q\int_{-\infty }^{+\infty }\left\vert \psi
(x=L,t)\right\vert ^{2}dt=-2\frac{q\eta }{v},  \label{deltaNq}
\end{equation}
where $\psi (x=L)$ was taken as per the unperturbed soliton (\ref{soliton}).
As concerns the Hamiltonian, in the absence of perturbations its
conservation follows from the continuity equation relating the Hamiltonian
density and current conjugate to it,
\[
\frac{\partial }{\partial t}\left[ \frac{1}{2}\left( \left\vert
\psi _{x}\right\vert ^{2}-|\psi |^{4}\right) \right]
=\frac{i}{4}\frac{\partial }{\partial x}\left[ \left( \psi
_{x}^{\ast }\psi _{xx}-\psi _{x}\psi _{xx}^{\ast }\right) +2|\psi
|^{2}\left( \psi \psi _{x}^{\ast }-\psi ^{\ast }\psi _{x}\right)
\right] .
\]Integration of this equation over the domain of $0<x<L$ gives rise to the
evolution equation for $H_{1}$ similar to its counterpart (\ref{E}) for the
norm,
\begin{equation}
\frac{dH_{1}}{dt}=q\left[ |\psi |^{4}+\frac{1}{4}\left( \psi _{xx}\psi
^{\ast }+\psi _{xx}^{\ast }\psi \right) \right] .  \label{dH/dt}
\end{equation}
Substituting unperturbed soliton (\ref{soliton}) in this relation and
subsequent integration yield the net change of $H_{1}$ induced by the bounce
of the soliton from the valve, cf. Eq. (\ref{deltaNq}):
\begin{equation}
\left( \Delta H_{1}\right) _{q}\equiv q\int_{-\infty }^{+\infty
}\frac{dH_{1}}{dt}dt=\frac{q\eta }{v}\left( \eta ^{2}-v^{2}\right)
.  \label{deltaHq}
\end{equation}

The change in the norm and energy incurred by the collision of the soliton
with the valve is to be compensated by matter which was pumped from the
reservoir and absorbed by the soliton sweeping the cavity. Assuming, in the
spirit of the perturbation theory, that the matter-wave background built up
in the cavity between passages of the soliton, $\psi _{0}(x)$, has a small
amplitude, and the soliton collects all the available matter, the
linearization of Eq. (\ref{psi}) yields
\begin{equation}
\psi _{0}(x)=i\kappa \phi _{0}(x)\tau (x),  \label{a(x)}
\end{equation}where $\tau (x)$ is the time between two passages of the soliton through a
given point, and
\begin{equation}
\phi _{0}(x)=\sqrt{\frac{2N_{2}}{L}}\sin \left( \frac{\pi x}{L}\right)
\label{gstate}
\end{equation}
is the ground-state wave function in the reservoir, with the number of atoms
$N_{2}$ in it; the depletion of the reservoir is neglected, which is a
natural assumption in the perturbative setting. As the soliton passes each
point twice in the course of the cycle of circulations, it is easy to see
that $\tau (x)$ has two branches,
\begin{equation}
\tau _{1}=2x/v,~\tau _{2}=2\left( L-x\right) /v.  \label{tau}
\end{equation}
Then, it follows from Eqs. (\ref{N}), (\ref{H}) and (\ref{tau}) that the
norm and energy collected by the soliton within the circulation cycle are
\begin{eqnarray}
\left( \Delta N_{1}\right) _{\kappa } &=&\left( \kappa \phi _{0}\right)
^{2}\int_{0}^{L}\left[ \tau _{1}^{2}(x)+\tau _{2}^{2}(x)\right] dx=
C_{N}\frac{\kappa ^{2}N_{2}L^{2}}{v^{2}},  \label{deltaNkappa} \\
\left( \Delta H_{1}\right) _{\kappa } &=&-\frac{1}{2}\left( \kappa \phi
_{0}\right) ^{4}\int_{0}^{L}\left[ \tau _{1}^{4}(x)+\tau _{2}^{4}(x)\right]
dx=-C_{H}\frac{\left( \kappa ^{2}N_{2}\right) ^{2}L^{3}}{v^{4}},
\label{deltaHkappa}
\end{eqnarray}
where $C_{N}\equiv 4\left( \frac{1}{3}-\frac{1}{\pi ^{2}}\right) \approx
\allowbreak 0.93$ and $C_{H}=\frac{24}{5}-\frac{30}{\pi
^{2}}+\frac{189}{4\pi ^{4}}\approx \allowbreak 2.25$.

The periodic circulations are provided by obvious balance
conditions, $\left( \Delta N_{1}\right) _{q}+\left( \Delta
N_{1}\right) _{\kappa }=0$ and $\left( \Delta H_{1}\right)
_{q}+\left( \Delta H_{1}\right) _{\kappa }=0$. Using Eqs.
(\ref{deltaNq}), (\ref{deltaHq}) and (\ref{deltaNkappa}),
(\ref{deltaHkappa}), these conditions determine equilibrium values
of the soliton's amplitude and velocity, $\eta _{0}$ and $v_{0}$:
\begin{eqnarray}
\eta _{0}^{2} &=&\frac{C_{N}}{2}\frac{\kappa ^{2}N_{2}L^{2}}{q},~
\label{eta} \\
v_{0} &=&\eta _{0}-\delta v_{0},~\delta v_{0}\equiv
\frac{C_{H}}{2}\frac{\left( \kappa ^{2}N_{2}\right)
^{2}L^{3}}{q\eta ^{5}}.  \label{v}
\end{eqnarray}
Within the framework of the analytical model it is also possible to analyze
stability of this equilibrium solution, by assuming small
deviations of $\eta $ and $v$ from values (\ref{eta}) and
(\ref{v}). For the corresponding nonequilibrium solutions, Eqs.
(\ref{deltaNq}), (\ref{deltaHq}) and (\ref{deltaNkappa}),
(\ref{deltaHkappa})\ give rise to a \textit{map}, $2\eta
\rightarrow 2\eta ^{\prime }=2\eta +\left( \Delta N_{1}\right)
_{q}+\left( \Delta N_{1}\right) _{\kappa }$, $H_{1}\rightarrow
H_{1}^{\prime }=H_{1}+\left( \Delta H_{1}\right) _{q}+\left(
\Delta H_{1}\right) _{\kappa } $ [in fact, it is more convenient
to analyze the map written in terms of $\eta $ and reduced
Hamiltonian, $h\equiv H_{1}-(2/3)\eta ^{3}$]. Then, equilibrium
solution (\ref{eta}) and (\ref{v}) is a fixed point (FP)\ of the
map, and its stability is determined by \textit{multiplicators}
$\mu $, i.e., eigenvalues of the map linearized around the FP.
Direct calculation yields two eigenvalues, $\mu _{\eta }=1-2\left(
q/\eta _{0}\right) $ and $\mu _{h}=1-\left( q/\eta _{0}\right) $.
As $q$ is a small positive parameter, we conclude that $0<\mu
_{\eta },\mu _{h}<1$, hence the FP is stable indeed.

Despite many approximations adopted in the analytical version of
the model, the prediction for the soliton's amplitude given by Eq.
(\ref{eta}) generally agrees with numerical results. For instance,
in the example of the periodic regime displayed in Figs.
\ref{fig1}(a), \ref{fig2}(a) and \ref{fig3}(a), the averaged
maximum density of the solitonic waveform is $\left\langle \left(
\left\vert \psi (x,t\right\vert ^{2}\right) _{\max }\right\rangle
\simeq 27$, while Eq. (\ref{eta}) predicts, for the same case,
$\eta _{0}^{2}\approx 25$.

The analytical model also helps to explain the presence of the lower border
of the stability region in Fig. \ref{fig4}. Indeed, stable circulations of
the soliton are possible if the release of matter through the outlet can
compensate the permanent inflow of matter from the reservoir. Adopting a
rough estimate according to which $\eta $, $v$ and $q$ are on the same order
of magnitude, which is borne out by numerical data in the general case, and
making use of Eqs. (\ref{deltaNq}) and (\ref{deltaNkappa}), one can put the
latter condition in the form of $q^{3}>q_{\min }^{3}\sim \kappa
^{2}N_{2}L^{2}$. For parameter values corresponding to Fig. \ref{fig4}, this
yields $q_{\min }\sim 7$, which generally agrees with the situation shown in
that figure.

One can also understand why the lower stability border in Fig.
\ref{fig4} is nearly flat (in fact, the explanation is quite a
general one, and is not predicated on the validity of the above
analytical model). Indeed, it follows from Eq. (\ref{phi}) that
nonlinear deformation of ground state (\ref{gstate}) is negligible
under condition $\varepsilon \ll \varepsilon _{\max }\equiv \pi
^{2}/\left( N_{2}L\right) $. In the case corresponding to Fig.
\ref{fig4}, with $N_{2}=1000$ and $L=1$, one obtains $\varepsilon
_{\max }\simeq 0.01$, which complies with what one can observe in
Fig. \ref{fig4}.

A limitation of the quantitative validity of results produced by
the analytical model can be seen from the calculation of the
maximum value of the matter-release rate: Eqs. (\ref{E}),
(\ref{soliton}) and (\ref{eta}) yield a perturbation-theory result
for it, $R_{\max }=\left( C_{N}/2\right) \kappa ^{2}N_{2}L^{2}$,
which does not depend on the outlet parameter, $q$. On the other
hand, Fig. \ref{fig6} shows that, in the range of $5<q<17.5$, the
numerical results yield an approximate dependence, $R_{\max }\sim
q^{-0.75}$. However, the perturbation theory does not really apply
in this range.

\section{Conclusion}

In this paper, we have investigated the model of the matter-wave laser
generating a periodic array of coherent pulses, which was recently
introduced in Ref. \cite{we}. While only a relatively primitive regime of
vibrations of a broad immobile lump in the cavity was found in the original
consideration of the model, here we have demonstrated that exploration of a
broader parameter region reveals stable lasing modes accounted for by
periodic or quasi-periodic circulations of a narrow stable soliton in the
cavity; the soliton periodically generates an outcoupling pulse, hitting the
valve, and then gets replenished, absorbing matter pumped from the
reservoir. The existence of the stable circulation regime, and some features
of the numerically found stability regions were explained in the framework
of a simplified analytical model, that treated the intra-cavity soliton as a
quasi-particle, and predicted the stable regime by means of balance
conditions for the number of atoms and energy in the cavity.

Getting back from the Gross-Pitaevskii equations in the normalized form of
Eqs. (\ref{psi}), (\ref{phi}) to physical units, and making use of typical
values of physical parameters for the $^{7}$Li condensate, it is easy to
conclude that the circulation regime found in this work easily provides for
the generation of regular pulses built of $\mathcal{N}=10^{3}-10^{4}$ atoms
each (the vibrational lasing mode investigated in Ref. \cite{we} gave rise
to $\mathcal{N}\ $smaller by a factor of $\sim 100$). With the scaled value
of the initial norm, $N_{2}^{(0)}=1000$, employed in the examples, the laser
would generate $\sim 100$ pulses before the reservoir depletion will becomes
tangible. However, the latter number may be readily made indefinitely large,
as indefinite increase of $N_{2}^{(0)}$ does not affect the stability of the
circulation regime.

Undoing the rescaling which cast the Gross-Pitaevskii equation in
the normalized form of Eq. (\ref{psi}), it is easy to see that,
for an experimentally relevant value of the cavity length, $L=100$
$\mu $m, the time unit in the normalized equation corresponds to
physical time $t_{0}=mL^{2}/\hbar \approx 1$ s (recall $m$ is the
atomic mass, i.e., $m\approx \allowbreak 1.2\times 10^{-26}$ kg
for $^{7}$Li). Therefore, according to Fig. \ref{fig6}, a typical
circulation period is $\simeq 0.2$ s in physical units. Lastly,
detailed consideration of numerical data shows that the typical
value of the duty cycle characterizing the quality of the
generated pulse array is $\simeq 30\%$ (the quality of
small-amplitude arrays found in Ref. \cite{we} was poorer, with
the duty cycle $\sim 50\%$, i.e., the pulses were more overlapped
than in the circulation regime).

The realization of the proposed matter-wave soliton laser by means of
available experimental techniques seems to be quite feasible. It may also be
possible to consider another variant of the proposed matter-wave soliton
laser, with outlets at both edges at the cavity. The so modified system may
generate a double stream of matter-wave solitons, which, however, needs
separate consideration.

\section*{Acknowledgement}

The work of B.A.M. was supported, in a part, by the Israel Science
Foundation through the Center-of-Excellence grant No. 8006/03.

\end{document}